\documentclass[aps,prb,twocolumn,amsmath,amssymb,floatfix]{revtex4}

\usepackage{graphicx}

\newcommand{\green}[2]{\ensuremath{\langle\!\langle #1;#2\rangle\!\rangle}}
\newcommand{\corr}[1]{\ensuremath{\langle #1 \rangle}}

\begin{document}


\title{Ferromagnetism and non-local correlations in the Hubbard model}

\author{S.~Henning}
 \email{henning@physik.hu-berlin.de}
\author{W.~Nolting}
\affiliation{Lehrstuhl Festk\"orpertheorie, Institut f\"ur Physik, Humboldt-Universit\"at zu Berlin, Newtonstrasse 15, 12489 Berlin, Germany} 

\date{\today}

\begin{abstract}
We study the possibility and stability of band-ferromagnetism in the single-band
Hubbard model for the simple cubic (SC) lattice. A non-local self-energy is
derived within a modified perturbation theory. Results for the spectral density
and quasiparticle density of states are shown with special attention to the
effects of $\mathbf{k}$-dependence. The importance of non-local correlations for
the fulfillment of the Mermin-Wagner theorem is our main result. A
phase digram showing regions of ferromagnetic order is calculated for the three
dimensional lattice. Besides, we show results for the optical conductivity
and prove that already the renormalized one-loop contribution to the
conductivity cancels the Drude peak exactly in case of a local self-energy which
is not anymore true for a non-local self-energy.
\end{abstract}

\pacs{Valid PACS appear here}
\maketitle
\section{Introduction}
Band-ferromagnetism is bound to the existence of permanent magnetic moments belonging 
to itinerant electrons in a partially filled conduction band \cite{MetMag}. 
Archetypical representatives are the classical $3d$ ferromagnets Fe, Co, Ni. 
The microscopic interpretation of band-ferromagnetism is one of the most fundamental 
and also most complicated many-particle problems in condensed matter physics. 
It is expected to be due to the interplay between ordinary, spin-independent 
Coulomb interaction (strong and strongly screened) and kinetic energy in the 
frame determined by the Pauli principle.

A minimal model for the investigation of band-ferromagnetism was proposed
independently by Hubbard\cite{Hubbard1}, Kanamori\cite{Kanamori} and Gutzwiller\cite{Gutzwiller}.
Despite its simple appearance, the Hubbard model Hamiltonian forms a highly
non-trivial many body problem that cannot be treated rigorously for the general
case.

A big step forward in the understanding of correlation effects in the Hubbard model was 
the invention of the \textit{'Dynamical Mean Field Theory'} (DMFT), which
becomes an exact theory in the limit of infinite lattice dimensions
\cite{MetznerVollhardt89, MuellerHartmann89, GeKoKrRo}. The DMFT maps the
lattice problem onto an effective single-impurity Anderson model (SIAM) which
can be solved numerically essentially exactly by use of e.g. Quantum Monte Carlo
methods \cite{Gull11}.

One shortcoming of the otherwise highly successful DMFT is the locality 
(wave-vector independence) of the electronic self-energy strictly valid only 
for $d=\infty$. So it may be questionable, e.g., whether such a self-energy is 
sufficient to describe angle-resolved photoemission results. Recent efforts have 
therefore been focussed on regaining a certain degree of non-locality in the
DMFT self-energy \cite{MaiJarPruHet,Held07,Sadovskii05,Rubtsov08,Fuchs11}.

There are other approaches to the non-locality of the self-energy at low dimensions
$d=2,3$. Coming from the weak coupling limit, Schweitzer and Czycholl proposed
a method for solving the highly involved wave vector summations that appear
already in second order diagrammatic perturbation theory \cite{SchweitzerCzycholl91}.
Kakehashi and Fulde used a projection operator method combined with the coherent
potential approximation for an investigation of the non-local excitation spectra
\cite{Kakehashi04}.

Concerning ferromagnetism the few exactly known results for the Hubbard model
are of great value and can be used as a test frame for approximate theories.
The Nagaoka theorem \cite{Nagaoka66} states, that a saturated ferromagnetic
order is the ground state for $U=\infty$ when one hole/electron is introduced
into the half filled band for the simple cubic (SC) lattice in three dimensions
(3D). The Mermin-Wagner theorem \cite{Mermin66} rules out ferromagnetic and
anti-ferromagnetic order in the Hubbard model in dimensions $d\le2$ for finite
temperatures [\onlinecite{Gosh84,Gelfert99,NoltingRamakanth09}]. For the infinite
dimensional SC and FCC lattice the existence of ferromagnetism was proved by
DMFT calculations [\onlinecite{Obermeier97,Zitzler02,Ulmke98,Vollhardt99}].

Apart from these rigorous results several works have investigated the possibility of
ferromagnetism in the Hubbard model within an approximation. DMFT calculations where done for the 
3D SC and FCC lattice [\onlinecite{Ulmke98,Schmitt11}] and the influence of
next-nearest-neighbour hopping was investigated in [\onlinecite{Peters09}].
Ferromagnetism in various lattices was investigated with a spectral density
approach (SDA) self-energy in [\onlinecite{Herrmann97,Herrmann297}]. Variational
methods have been used in [\onlinecite{Hanisch97,Fazekas90,Guenther11}].

A general trend can be read from these calculations. Two main ingredients favor
ferromagnetism in the Hubbard model. An asymmetric density of states (DOS)
(e.g. the FCC DOS) and non-bipartite lattices with frustration in
the anti-ferromagnetic correlations, which can be generated by introducing
next-nearest neighbor hopping $t'$. This shows the competitive character of
ferro- and anti-ferromagnetic correlations in the Hubbard model.

In this paper we investigate the influence of non-local correlations on
ferromagnetic order in the SC lattice. To this end we shall apply the \textit{Modified
Perturbation Theory} (MPT), which was originally used only for solving the SIAM 
within the DMFT procedure, directly to the full Hubbard problem. The MPT leads to 
an explicitly wave-vector dependent self-energy which decisively determines the single-electron 
spectral density. It is well known that the latter provides the bare line shape 
of a spin and angle-resolved (direct or inverse) photoemission experiment. 

The paper is organized as follows. In section \ref{sec:theory} we introduce the
Hubbard model Hamiltonian, derive the MPT self-energy and discuss its
properties. Then thermodynamic quantities as the paramagnetic static
susceptibility and the optical conductivity are derived.\\
In section \ref{sec:compmeth} the numerical methods for dealing with the
complicated momentum summations are presented. \\
Section \ref{sec:results} contains the results and interpretation of our numerical 
calculations. Finally we give a summary and conclusion in section
\ref{sec:summary}. 

\section{\label{sec:theory}theory}
\subsection{Hubbard model}
The Hamiltonian of the Hubbard model is given by:
\begin{equation}
    H =  t\sum_{\langle i,j\rangle\sigma}c^+_{i\sigma}c_{j\sigma}
        +\sum_{i\sigma}(z_{\sigma}B+t_0)\hat{n}_{i\sigma}
        +U\sum_{i}\hat{n}_{i,\uparrow}\hat{n}_{i,\downarrow}.
\label{eq:hamiltonian}
\end{equation}
Here $t$ denotes the nearest neighbor hopping strength (the sum over
$\mathbf{R}_j$ extends only over the nearest neighbors of $\mathbf{R}_i$), $U$ is the local
coulomb repulsion and $B$ an homogeneous external magnetic field ($z_{\uparrow
\downarrow}=\pm 1$). We have chosen the band center of gravity $t_0=0$ and
the hopping strength $t$ such that the free electronic bandwidth $W$ is
equal to one throughout the paper. An (approximate) solution of the
Hubbard model is found, if we are able to calculate the electronic Green's
function (GF):
\begin{equation}
G_{\mathbf{k}\sigma}(E) =
    \left(E+\mu-\epsilon(\mathbf{k})-\Sigma_{\mathbf{k}\sigma}(E)\right)^{-1}
\label{eq:elect_green}
\end{equation}
or more precisely the electronic self-energy $\Sigma_{\mathbf{k}\sigma}(E)$. 
\subsection{self-energy}
It is now a well-known fact, that the self-energy of the Hubbard model becomes a
purely local quantity in the limit of infinite dimensions ($d\rightarrow \infty$)
\cite{MetznerVollhardt89, MuellerHartmann89}. However the $\mathbf{k}$-dependence 
will certainly play a crucial role in the more realistic case of
$d=2,3$. In the weak coupling limit ($U \ll W$) the second order perturbation theory
(SOPT) is a good starting point for the investigation of non local correlation
effects. The SOPT self-energy is given by \cite{SchweitzerCzycholl91}:
\begin{widetext}
\begin{eqnarray}
\label{eq:sopt_self}
\lefteqn{\Sigma_{\mathbf{k}\sigma}(E) =
    \Sigma_{\sigma}^{(HF)}+\Sigma_{\mathbf{k}\sigma}^{(SOC)}(E)}  \\
    & = & U\corr{n_{-\sigma}}   
    + U^2\sum_{\mathbf{R}}\mathrm{e}^{i\mathbf{kR}}
        \int\mathrm{d}xS_{\mathbf{R}\sigma}(x)
        \int\mathrm{d}yS_{\mathbf{R}-\sigma}(y)
        \int\mathrm{d}zS_{\mathbf{-R}-\sigma}(z)
        \frac{f_{-}(x)f_{-}(y)f_{-}(-z)+f_{-}(-x)f_{-}(-y)f_{-}(z)}
             {E+i0^{+}-x-y+z}. \nonumber
\end{eqnarray}
\end{widetext}
The sum extends over all lattice sites $\mathbf{R}$. Schweitzer and Czycholl
\cite{SchweitzerCzycholl91} gave a method for the calculation of
(\ref{eq:sopt_self}) by collecting all symmetry equivalent points in shells and
recast the sum over lattice sites into a sum over shells. They showed, that this
sum can be truncated after a finite number of shells. However, their method of
calculating the real and imaginary part of (\ref{eq:sopt_self}) is still
numerically very demanding. We will show in section \ref{sec:compmeth} how to
speed up the computation to allow fully self consistent calculations for
arbitrary band fillings $n$. 
There is a certain arbitrariness in (\ref{eq:sopt_self}) concerning the
spectral densities (SD) appearing in the formula. In an expansion strictly to
order $U^2$ the free SD has to be chosen \footnote{This statement is not fully
accurate since the first (Hartree) term contains a partial sum of
diagrams of all orders already, when one chooses the full self-consistent $\langle n_{-\sigma}
\rangle$ as we are doing in this work.}. But one could also renormalize the
theory by using the full SD in a self-consistent manner. It turns out, that only
the first choice will reproduce certain exact results \cite{Vollhardt92}. 
To be specific we give the form of the SD used:
\begin{equation}
S_{\mathbf{R}\sigma}^{(0)}(x) =
    \frac{1}{N}\sum_{\mathbf{k}}\mathrm{e}^{i\mathbf{kR}}
    \delta(x+\mu_{\sigma}^{(0)}-\epsilon(\mathbf{k})).
\label{eq:SD0}
\end{equation}
$\mu_{\sigma}^{(0)}$ is fixed by the condition, that the free occupation number is
equal to the full occupation: $\corr{n_{\sigma}}^{(0)} = \corr{n_{\sigma}}$. 
Notice, that this choice of $\mu_{\sigma}^{(0)}$ is 
equivalent to a SOPT ``around Hartree-Fock (HF)'' at half filling where the full
$\mu$ from (\ref{eq:elect_green}) is taken in the HF-SD.
Only this choice will result in the now widely accepted three peak structure of
the density of states (DOS) and will give a ``smooth'' change of the DOS away
from half filling. For more qualitative discussions we refer the reader to the
results section.\\
To extend the validity of the self-energy to larger values of $U$ we use the
following Ansatz for a modified perturbation theory (MPT):
\begin{equation}
    \Sigma_{\mathbf{k}\sigma}(E) = U\corr{n_{-\sigma}} +
    \frac{a_{\mathbf{k}\sigma}\Sigma_{\mathbf{k}\sigma}^{(SOC)}(E)}
         {1-b_{\mathbf{k}\sigma}\Sigma_{\mathbf{k}\sigma}^{(SOC)}(E)}.
\label{eq:ansatz_mpt}
\end{equation}
This form for the self-energy was proposed by Kajueter and
Kotliar\cite{KajueterKotliar96} for the Anderson impurity model (SIAM). They used
the first two spectral moments and an additional condition for the chemical
potential to fix the parameters $a_{\mathbf{k}\sigma}$ and $b_{\mathbf{k}\sigma}$.
This method of fixing the parameters was afterwards 
modified by Potthoff, Wegner and Nolting \cite{PotthoffWN97} for the same model
in order to reproduce the first four moments of the SD correctly. We will follow
this latter approach but now applied to the full lattice Hamiltonian
(\ref{eq:hamiltonian}). \\
To fix the appearing constants in (\ref{eq:ansatz_mpt}) we use the high energy
expansion of the self-energy for the Hubbard model:
\begin{equation}
\Sigma_{\mathbf{k}\sigma}(E) =
\sum_{m=0}^{\infty}\frac{C_{\mathbf{k}\sigma}^{(m)}}{E^m}.
\label{eq:expand_self}
\end{equation}
The first three coefficients can be obtained from the first four moments
$M_{\mathbf{k}\sigma}^{(m)}$ of the SD\cite{NoltingRamakanth09} 
via the high energy expansion of the electronic GF (\ref{eq:elect_green}):
\begin{equation}
G_{\mathbf{k}\sigma}(E) =
\frac{1}{E}\sum_{m=0}^{\infty}\frac{M_{\mathbf{k}\sigma}^{(m)}}{E^m}
\label{eq:expand_green}
\end{equation}
and are given in the appendix (\ref{eq:coeff_self1}).
By expanding also the rhs of (\ref{eq:ansatz_mpt}) we can determine the
coefficients $a_{\mathbf{k}\sigma}, b_{\mathbf{k}\sigma}$ and get finally the MPT
self-energy:
\begin{equation}
    \Sigma_{\mathbf{k}\sigma}(E) =
    U\corr{n_{-\sigma}}+\left((\Sigma_{\mathbf{k}\sigma}^{(SOC)}(E))^{-1} + 
    \frac{D_{\mathbf{k}\sigma}^{(2)}-C_{\mathbf{k}\sigma}^{(2)}}
         {(C_{\mathbf{k}\sigma}^{(1)})^2}\right)^{-1},
\label{eq:self_mpt}
\end{equation}
where $D_{\mathbf{k}\sigma}^{(2)}$ denotes the third moment of
(\ref{eq:sopt_self}) as given in the appendix (\ref{eq:coeff_self2}).\\
The MPT self-energy can be proved to be exact in a variety of limiting cases.
It trivially fulfills the limits of $U=0$ and $n=0, n=2$. More interesting is
case of zero bandwidth limit $t\rightarrow 0$. 
A straightforward calculation yields:
\begin{equation}
\Sigma_{\mathbf{k}\sigma}^{(W\rightarrow0)}(E)=U\corr{n_{-\sigma}}
    + \frac{U^2\corr{n_{-\sigma}}(1-\corr{n_{-\sigma}})}
           {E+z_{\sigma}B+\mu-U(1-\corr{n_{-\sigma}})},
\label{eq:zero_bw_limit}
\end{equation}
which is indeed the correct form of the ``atomic limit''
\cite{NoltingRamakanth09}. Expanding the self-energy for small $U$ reproduces
the result of perturbation theory (\ref{eq:sopt_self}) with corrections only of 
order $U^3$. Therefore the MPT should be correct for small U and show e.g.
Fermi liquid behavior \cite{SchweitzerCzycholl91}. \\
Since the first four spectral moments are reproduced correctly by construction, 
they should be correct at large energies $|E|\gg1$, too. In particular the position and
shape of the upper (lower) Hubbard band for $n<1$ ($n>1$) will become exact in
the strong coupling limit ($U\gg1$) which is known to be a weak point of SOPT
alone. In this respect the MPT self-energy is in accordance to the $t/U$
strong coupling expansion of Harris and Lange \cite{Harris67, Potthoff98}.
To conclude this discussion we summarize our findings. We have proposed a
fully $\mathbf{k}$-dependent MPT self-energy which fulfills the atomic limit
and shows reasonable behavior in the weak and strong coupling region. Therefore
there is well-founded hope that our theory will give reasonable results also in
intermediate coupling region.  
\subsection{thermodynamics and transport}
\subsubsection{paramagnetic static susceptibility}
The developed theory allows for a self-consistent calculation of the magnetization in
a possible appearing ferromagnetic region. To test the system regarding a ferromagnetic
phase transition, we will calculate the paramagnetic static susceptibility, which
is defined as follows:
\begin{equation}
\hat{\chi}^{(p)}(T)=\sum_{\sigma}\partial_{B}(z_{\sigma}\corr{n_{\sigma}})|_{T,B=0,\corr{n_{\uparrow}}=\corr{n_{\downarrow}}}.
\label{eq:para_susz}
\end{equation}
The zero crossings of the inverse of (\ref{eq:para_susz}) indicate the points 
where the paramagnetic phase become susceptible to a ferromagnetic phase
transition.

For an evaluation of (\ref{eq:para_susz}) one has to perform the
derivative analytically and get after a lengthy calculation an explicit form for
the susceptibility as a functional of the (self-consistently determined) 
paramagnetic self-energy. Since the expressions are rather long, we do not give
them here.

\subsubsection{optical conductivity}
\label{sssec:opt_cond}
The optical conductivity in linear response is given by
the retarded current-polarization GF \cite{Nolting09}:
\begin{equation}
\sigma^{\beta\alpha}(E)=-\green{\hat{j}^{\beta}}{\hat{P}^{\alpha}}_{E},
\label{eq:cond_1}
\end{equation} where $\alpha,\beta$ denote the Cartesian coordinates of the
operators. By writing down the EQM of this GF and exploiting the connection
$\hat{\mathbf{j}}=-i\frac{1}{N}[\hat{\mathbf{P}},H]_{-}$ this can be rewritten
as:
\begin{equation}
\sigma^{\beta\alpha}(E)=-\frac{\corr{[\hat{j}^{\beta},\hat{P}^{\alpha}]_{-}}}{E}
    +iN\frac{\green{\hat{j}^{\beta}}{\hat{j}^{\alpha}}}{E}.
\label{eq:cond_2}
\end{equation}
For a tight binding (nearest neighbor hopping) model the operators are given as  
$\hat{\mathbf{P}}=q\sum_{i,\sigma}\mathbf{R}_{i}\hat{n}_{i\sigma}$ and
$\hat{\mathbf{j}}=-\frac{iq}{N}t\sum_{\langle im\rangle,\sigma}(\mathbf{R}_{i}-\mathbf{R}_{m})c_{i\sigma}^{+}c_{m\sigma}$. 
With these operators the first term of (\ref{eq:cond_2}) can be calculated and
in case of a simple cubic lattice simplified to give the zero frequency Drude 
weight of conductivity:
\begin{equation}
\mathrm{Re}(\sigma_{D}^{\beta\alpha}(E+i0^+)) =
    -\pi\delta_{\alpha\beta}\delta(E)\frac{2tq^2}{N}\sum_{\mathbf{k}\sigma}\cos{(k_{\alpha})}
    \corr{\hat{n}_{\mathbf{k}\sigma}}.
\label{eq:drude_cond}
\end{equation}
The second term in (\ref{eq:cond_2}) is the current-current GF. It represents
the influence of electronic correlations. We approximate this GF on the 'one
loop' level in a diagrammatic expansion. The explicit calculation is given in
appendix (\ref{sec:optcondderiv}) and it is shown that the real part
consists of two parts. One is proportional to $\delta(E)$ and cancels the Drude
peak in case of a local self-energy exactly. The second term yields:
\begin{eqnarray}
\lefteqn{\mathrm{Re}(\sigma_{C}^{\beta\alpha}(E+i0^+)) = \delta_{\alpha\beta}\frac{\pi q^2}{N}
        \sum_{\mathbf{k}\sigma}v_{\mathbf{k}\alpha}v_{\mathbf{k}\beta}}
        \nonumber \\
    & & \int\mathrm{d}x S_{\mathbf{k\sigma}}(E+x)S_{\mathbf{k\sigma}}(x)
        \frac{f_{-}(x)-f_{-}(E+x)}{E}, 
\label{eq:jj_cond}
\end{eqnarray}
with $v_{\mathbf{k}\alpha}=\partial_{k_{\alpha}}\epsilon_{\mathbf{k}}$. 

Note, that this approximation becomes a rigorous result in infinite dimensions because
the self-energy is a local quantity and all higher vertex corrections vanish 
in this case \cite{PruschkeCox93}.

We will only show results for the contribution from (\ref{eq:jj_cond}) hoping
that the exact cancellation is retained in the $\mathbf{k}$-dependent case approximately
at least. The neglect of vertex corrections can not be justified rigorously in case
of a $\mathbf{k}$-dependent self-energy, because Ward identities may be
violated. We show therefore only results for the 3D SC lattice, where the
$\mathbf{k}$-dependence of the self-energy is not so pronounced (especially at
the Fermi level) and refer the reader to the more specialized literature for a
thorough discussion of this point \cite{Bergeron11}.  
\section{\label{sec:compmeth}computational methods}
The self-energy (\ref{eq:self_mpt}) is a functional of the chemical potential and different correlation
functions. It has to be calculated self-consistently. To this end we need a fast way of calculating
integrals of the form:
\begin{equation}
\corr{\dots}=\frac{1}{N}\sum_{\mathbf{k}}\int\mathrm{d}Ef_{-}(E)F_{\mathbf{k}}(E)S_{\mathbf{k}}(E),
\label{eq:k_E_int}
\end{equation}
where $f_{-}(E)$ is the Fermi function, $F_{\mathbf{k}}(E)$ a polynomial of low order in $E$ and 
$S_{\mathbf{k}}(E)$ the full SD. It is hopeless to perform the four dimensional integral directly
because the SD is a strongly peaked function. However one can replace the energy integration by
a sum over the poles of $f_{-}(E)$ in the upper complex plane. The usual Matsubara form of the
Laurent expansion of $f_{-}(E)$ is not suitable here, because it converges very
slowly with an increasing number of poles. 
Recently Ozaki [\onlinecite{Ozaki07}] proposed a different pole expansion
for the Fermi function which gives a good approximation for a large energy domain down to very
low temperatures with only a few hundred poles. We use this expansion for a numerical very accurate
determination of the energy integral. The remaining $\mathbf{k}$ integration over the irreducible
wedge of the Brillouin zone can then be performed directly.\\
For the exposed procedure we need the values of the SOPT self-energy at the Ozaki poles 
in the complex plane. For their determination we first rewrite (\ref{eq:sopt_self}) as a 
sum over shells of symmetry equivalent points:
\begin{equation}
\Sigma_{\mathbf{k}\sigma}^{(SOC)}(E) = U^2\sum_{s=0}^{N}G_{\mathbf{k}}^{(s)}\Sigma_{\sigma}^{(s)}(E), 
\label{eq:soc_shellsum}
\end{equation}
where $G_{\mathbf{k}}^{(s)}$ denotes the $\mathbf{k}$ dependent shellfactor given by the sum over
the exponentials in (\ref{eq:sopt_self}) within one shell and $\Sigma_{\sigma}^{(s)}(E)$ the remaining
energy dependent part. The imaginary part of the latter is given by:
\begin{eqnarray}
\lefteqn{\mathrm{Im}\Sigma_{\sigma}^{(s)}(E)=-\pi\int\mathrm{d}x\int\mathrm{d}y} \\
& &	S_{s\sigma}^{(0)}(x)S_{s-\sigma}^{(0)}(y)
	S_{s-\sigma}^{(0)}(x+y-E)F(x,y,x+y-E), \nonumber
\label{eq:imagsopt}
\end{eqnarray}
where $F(x,y,z)$ denotes the product of Fermi functions in (\ref{eq:sopt_self}).
This twofold convolution can be solved very efficiently by a fast Fourier transform. 
From this $\Sigma_{\sigma}^{(s)}(E)$ can be obtained in the complex plane via the spectral
representation of Greens functions \cite{Nolting09}.
The free shell density of states needed for the calculation we have computed and
stored beforehand to very high precision. The number of shells necessary to get
converged results depends strongly on the coordination number of the underlying
lattice. After some tests we have used 201 shells for the 2D and 61 shells for
the 3D lattice for all calculations in this work, which was sufficient to get
mostly converged self-energies. 

\section{\label{sec:results}results and discussion}
Our $\mathbf{k}$-dependent self-energy allows the description of homogeneous 
phases (paramagnetic/ferromagnetic). 
We will discuss the electronic properties in the paramagnetic state of the three dimensional
 Hubbard model first, knowing well that in certain parameter regimes (e.g. near half filling
at low temperatures) anti-ferromagnetism is expected in principle. This restriction
is shared with other self-energy approaches (e.g. DMFT calculations) and a comparison to these
should render our results useful. 

\subsection{QDOS and spectral density}
\begin{figure}
    \includegraphics[width=0.9\linewidth]{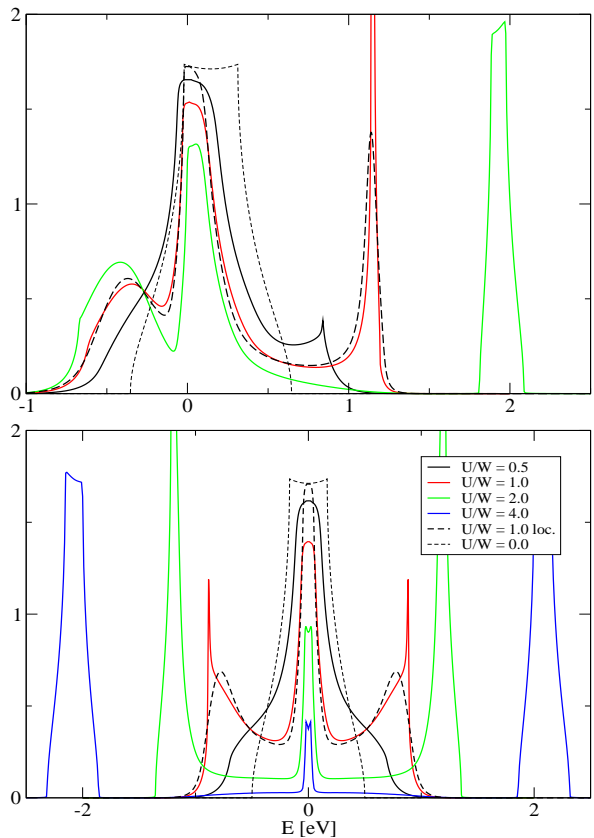}
    \caption{\label{fig:qdos}(color online) QDOS for various interaction
    parameters $U/W$; full lines: $\mathbf{k}$-dependent self-energy results;
    broken lines: local self-energy and interaction free result; Parameters:
    $T=10\mathrm{K}$, upper panel:$n=0.75$, lower panel:$n=1.0$}
\end{figure}

\begin{figure}
    \includegraphics[width=0.9\linewidth]{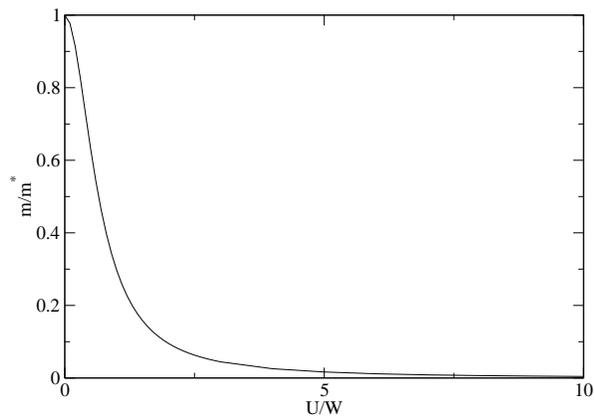}
    \caption{\label{fig:invmass}Inverse effective mass as function of
    interaction parameter $U/W$; Parameters: $T=10\mathrm{K}$, $n=1.0$}
\end{figure}

\begin{figure}
    \includegraphics[width=0.98\linewidth]{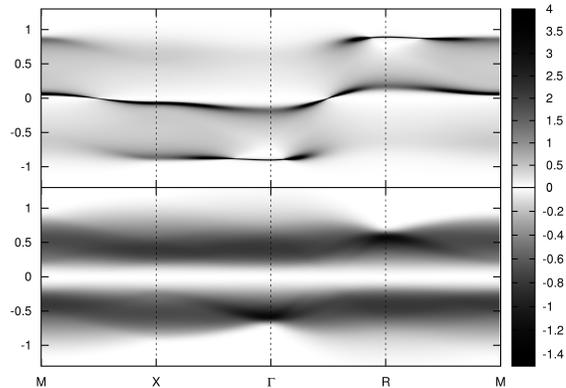}
    \caption{\label{fig:sdself3d}Spectral density (upper figure) and imaginary
    part of MPT self-energy (lower figure) along special directions in the first 
    Brillouin zone of the simple cubic 3D lattice:
    $M(\pi,\pi,0)\rightarrow X(\pi,0,0)\rightarrow \Gamma(0,0,0)\rightarrow
    R(\pi,\pi,\pi) \rightarrow M$; Parameters: $n=1$, $T=10\mathrm{K}$, $U/W=1$}
\end{figure}

\begin{figure}
    \includegraphics[width=0.9\linewidth]{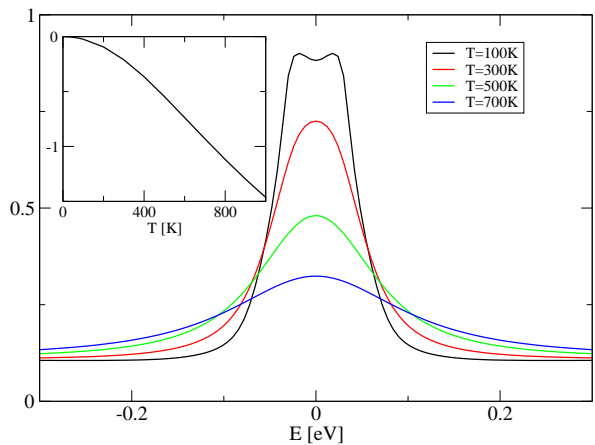}
    \caption{\label{fig:dos_T}(color online) QDOS (Kondo resonance) at various temperatures
    $T$; Inset: averaged imaginary part of MPT self-energy at Fermi level $E=0$;
    Parameters: $n=1$, $U/W=2$.}
\end{figure}

In Fig.~\ref{fig:qdos} we show the quasiparticle density of states (QDOS)
at low temperature ($T=10\mathrm{K}$) for two different band fillings and various
interaction parameters $U$.
The three peak structure of the QDOS is clearly visible. Upper and lower Hubbard band are 
roughly separated by the coulomb interaction strength $U$ and there appears a Kondo
resonance near the Fermi level ($E=0$ eV in figures). By increasing 
$U$, this resonance decreases but stays finite also for large $U$. As a consequence
there is no clear metal-insulator transition (MIT) for $n=1$ as is found in DMFT calculations.
To illustrate this further we show the inverse effective mass: 
$\frac{m}{m^{*}}=(\frac{1}{N}\sum_{\mathbf{k}}(1-\Re\Sigma_{\mathbf{k}}^{'}(0))^{-1}$
in Fig.~\ref{fig:invmass}. 
Although there is no clear transition point the system should be insulating above
$U/W\approx10$ due to the large effective mass of the quasiparticles. This finding is
in qualitative agreement with the non-local theory of Kakehashi and Fulde \cite{Kakehashi04}.\\
The comparison of the local approximation (only zeroth shell of the SOC is taken into account)
and the full $\mathbf{k}$-dependent calculation shows decisive effects of the latter.
Whereas the local theory fulfills the Luttinger theorem (the QDOS at the Fermi level
is equal to the free DOS), the $\mathbf{k}$-dependence leads to a reduction of states at this point.
This is understandable because the Luttinger-Ward argument only holds for local self-energies 
\cite{MuellerHartmann89_2,Luttinger60}. Another effect of the $k$-dependence are the
peaks in the upper ($n=0.75,1.0$) and lower ($n=1.0$) Hubbard band for intermediate
coupling ($U/W \approx 1$). For interpretation of these peaks we have plotted the
spectral density together with the imaginary part of the self-energy along special
directions within the first Brillouin zone for $n=1$ in Fig.~\ref{fig:sdself3d}. 
The self-energy shows typical Fermi liquid behavior.
The imaginary part is near zero (zero only at $T=0$) at the Fermi level and decreases 
quadratically with increasing energy. This leads to increasing damping effects in
the dispersion of the Kondo resonance particularly strong at the $\Gamma$ and $\mathrm{R}$
points. At these points the self-energy shows a strong enhancement around
$E=-/+0.5$ respectively and increases then abruptly to zero for lower/higher energies. 
Therefore we find no damping effects
in this energy region and quasiparticle states with energies lower/higher than the
threshold energy will have infinite lifetimes. This is clearly visible in the spectral
density where in the lower/upper Hubbard band "bridges" of sharply peaked states appear
at both $\mathbf{k}$ points. Similar effects where found in SOPT calculations by
Schweitzer and Czycholl\cite{SchweitzerCzycholl91} and in the projection operator method 
of Kakehashi and Fulde\cite{Kakehashi04} but here they are much more pronounced.
The reason is most likely the increased
numerical accuracy of our calculation (see section \ref{sec:compmeth}).\\
Comparing our results to a recent \textit{dynamical cluster
approximation}\cite{Fuchs11} (a cluster extension of the DMFT in order to retain
a $\mathbf{k}$-dependence) there is good
agreement below the MIT ($U/W=0.67$ in the mentioned
work). The dispersion of the Kondo resonance and the maxima of the lower/upper
Hubbard bands show essentially the same behavior as in our calculation. Above
the MIT ($U/W=1$) there are of course discrepancies because of the missing MIT
in the MPT.\\
With increasing temperature damping effects will become more important. This is
shown in Fig.~\ref{fig:dos_T}. The Kondo resonance peak is diminished with raising
temperature and tend to vanish completely at higher temperatures.
 The reason for
this can be found in the inset of Fig.~\ref{fig:dos_T}. With increasing temperature the averaged
imaginary part of the self-energy at the Fermi level decreases starting from
zero at $T=0K$. For low temperatures a typical Fermi liquid behavior
is obtained ($\sim T^2$).

\begin{figure}
    \includegraphics[width=0.98\linewidth]{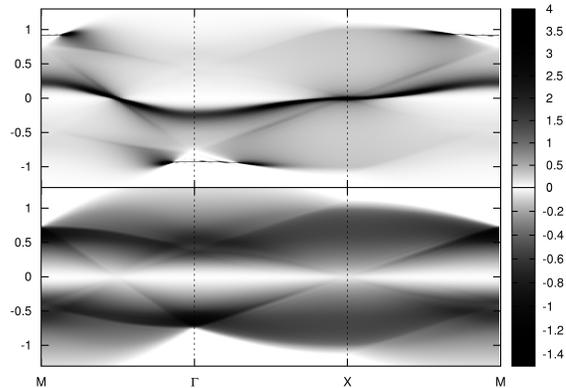}
    \caption{\label{fig:sdself2d}Spectral density (upper figure) and imaginary
    part of MPT self-energy (lower figure) along special directions in the 
    first Brillouin zone of the simple cubic 2D lattice:
    $M(\pi,\pi)\rightarrow \Gamma(0,0)\rightarrow X(\pi,0) \rightarrow M$; 
    Parameters: $n=1$, $T=10\mathrm{K}$, $U/W=1$}
\end{figure}

\begin{figure}
    \includegraphics[width=0.9\linewidth]{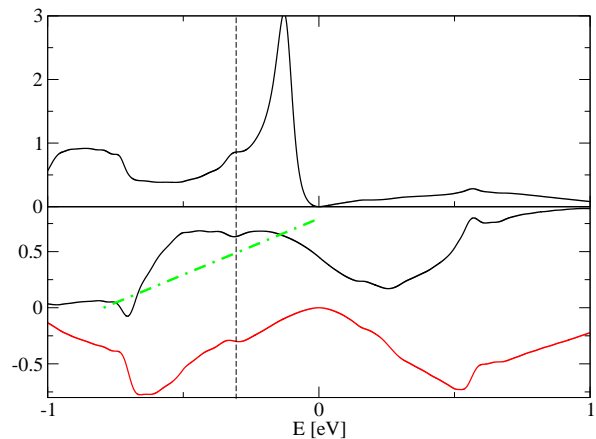}
    \caption{\label{fig:shadowexp}(color online) Spectral density (upper figure)
    and real part (lower figure: black line) and imaginary part (red line) of
    the self-energy for the simple cubic 2D lattice at 
    $\mathbf{k}=(\frac{6\pi}{20},\frac{6\pi}{20})$. Vertical broken line:
    position of the shadow band. The crossing point the green dash-dotted line
    with the real part of the self-energy marks the position of quasiparticle
    excitations. Parameters: $n=1$, $U/W=1$, $T=10\mathrm{K}$.}
\end{figure}

For the two dimensional SC lattice the effects of a non-local self-energy should
be more drastic then in 3D as a direct consequence of the reduced coordination
number. Fig.~\ref{fig:sdself2d} shows the spectral density and imaginary part of the
self-energy of the SC 2D lattice at half filling. We find again states with 
infinite lifetime at $\Gamma$ and $M$ point due to the vanishing imaginary part
of the self-energy. The self-energy shows Fermi liquid behavior ($\sim E^2$) at
large portions of the Brillouin zone. However, this behavior is changed near
the Fermi surface. At the $X$ point and the midpoint between $\Gamma$ and $M$
($\frac{\pi}{2},\frac{\pi}{2}$) the self-energy shows a linear energy dependence.
This ``marginal Fermi liquid'' behavior is a direct consequence of the perfect nesting
properties ($\epsilon(\mathbf{k})\approx\epsilon(\mathbf{k+Q})$, where
$\mathbf{Q}$ is a reciprocal lattice vector) 
of the 2D Fermi surface \cite{Hirsch84,Virosztek90} at half filling.
Electrons can be scattered efficiently due to the large phase space. 
This leads to a second effect in the spectral density - the
formation of ``shadow'' bands. They are visible as dark lines running from the
special points $X$, ($\frac{\pi}{2},\frac{\pi}{2}$) to $\Gamma$, $M$ with
a slope determined by the condition $E\approx \epsilon(\mathbf{k+Q})$.  
These shadow features, which have already been described by Vilk \cite{Vilk96}, are 
no real quasi particle band but merely thermal excitations corresponding to a 
local minimum of the imaginary part of the self-energy. To illustrate this point
we have plotted the spectral density (upper panel) and self-energy (lower panel)
at $\mathbf{k}=(\frac{6\pi}{20},\frac{6\pi}{20})$ in Fig.~\ref{fig:shadowexp}. The green
dash-dotted line obeys $E+\mu-\epsilon(\mathbf{k})$ and its crossing points with
the real part of the SE (black line) define the quasiparticle excitations of the system. 
In the SD only one of the three excitations forms a peak (near $E=0$), the
others are strongly damped by a large imaginary part of the SE (red line). The
black dashed line marks the position of the shadow band. There is
no real excitation energy but we find a local minimum of the imaginary part of
the SE which leads to observed shadow feature.

\subsection{conductivity}
\begin{figure}
    \includegraphics[width=0.9\linewidth]{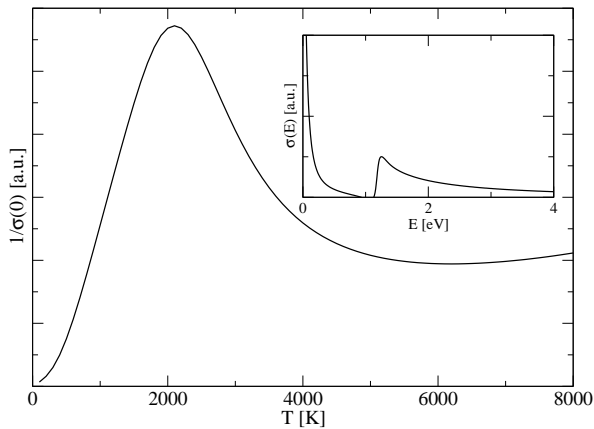}
    \caption{\label{fig:invleit}Correlation part of the resistivity (inverse of
    the static conductivity) as a function of temperature $T$ for the 3D simple
    cubic lattice. Inset: optical conductivity as function of $E$. 
    Parameters: $n=1$, $U=2$; Inset: $T=100\mathrm{K}$.}
\end{figure}

From (\ref{eq:jj_cond}) it becomes clear, that the optical conductivity is
mainly determined by the number of available quasiparticle states. In
Fig.~\ref{fig:invleit} the inverse static conductivity (resistivity) at half filling is shown as a function
of temperature for $U=2$. At low temperatures the resistivity increases
quadratically. This results from the reduction of the QDOS at Fermi level as
shown in Fig.~\ref{fig:dos_T}. The resistivity rises until the thermal energy is 
sufficient to excite electrons from the lower to the upper Hubbard band
($k_BT\approx U$). Then it will decrease going through a minimum and rise again.
The inset shows the optical conductivity at fixed temperature $T=100K$. The
conductivity decreases with increasing energy due to the lack of states between
the Kondo resonance and the upper Hubbard band. As soon as the energy is
sufficient to excite electrons from the Fermi level to the upper band, the 
conductivity will rise strongly reaching a maximum soon and decrease again.
These findings for the conductivity are in well agreement with DMFT results
for $U$ below the metal-insulator transition \cite{PruschkeCox93}

\subsection{inverse paramagnetic susceptibility and ferromagnetic phase
transition}
\begin{figure}
    \includegraphics[width=0.9\linewidth]{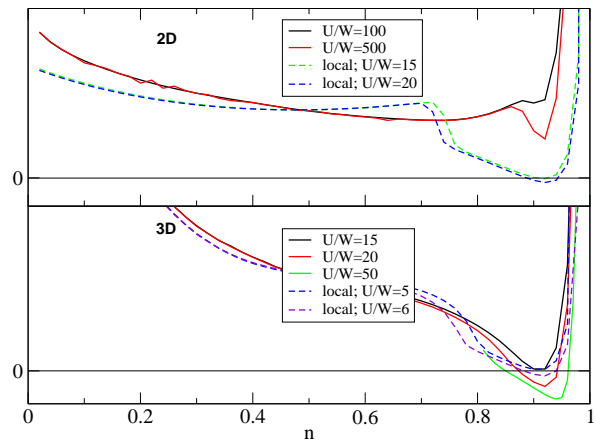}
    \caption{\label{fig:invsusz}Inverse paramagnetic susceptibility as a
    function of band filling $n$ for the 2D
    (upper figure) and 3D (lower figure) system calculated with the full
    $\mathbf{k}$-dependent self-energy. For the 2D system the result of local
    self-energy is shown also (broken lines). Parameters: $T=10\mathrm{K}$.}
\end{figure}

The inverse paramagnetic static susceptibility (IPS) (\ref{eq:para_susz}) can be used
as a tool for finding borders of a ferromagnetic phase transition in the $n$-$U$
diagram. Its zero crossings will mark the critical points. In
Fig.~\ref{fig:invsusz} we show the IPS for the 2D and 3D SC lattice at low
temperature ($T=10\mathrm{K}$) for various $U$. 
Whereas we find zero crossings above a critical $U/W\approx15$ in the 3D case,
there is no point of phase transition in the 2D case when we use the full non local
MPT self-energy. Only when using the truncated local version of the MPT we get a
phase transition in 2D also. This shows first of all, that the predictions of the non local
theory are in accordance with the Mermin Wagner theorem and secondly, that the
non-locality of the self-energy is crucial in oder to get the result. We would
like to mention that the IPS curve for $U/W=500$ is saturated in the sense, that increasing
$U/W$ further will not change its shape drastically.\\ We come now to the
discussion of the magnetic properties of the 3D system. The rather high critical
$U$ reflects the fact, that the SC lattice is not particularly susceptible to
ferromagnetism (the competing anti-ferromagnetism is not suppressed by
frustration, like in the FCC lattice).  
\begin{figure}
    \includegraphics[width=0.9\linewidth]{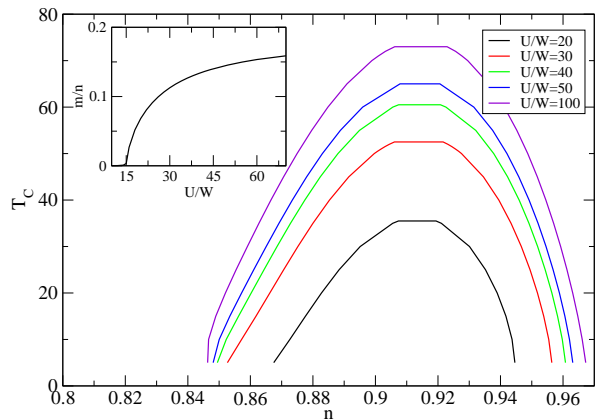}
    \caption{\label{fig:Tc}Curie temperature $T_C$ as function of band filling
    $n$ for various $U/W$. Inset: relative magnetization
    $(n_{\uparrow}-n_{\downarrow})/n$ as function of interaction strength $U/W$.
    Parameters inset: $n=0.9$, $T=10\mathrm{K}$.}
\end{figure}
The IPS crosses the zero axis at two points. These mark the lower and upper
bound of the ferromagnetic region. The Curie temperatures for different $U/W$
are plotted in Fig.~\ref{fig:Tc}. The maximal $T_C$ is reached for
a band filling of $n\approx0.91$. Starting from $U_{crit}/W\approx15$, $T_C$
increases quickly with increasing $U/W$ and running into saturation for larger
values of $U/W$. The same is true for the magnetization
$m=n_{\uparrow}-n_{\downarrow}$ and the phase border. The magnetization at
$n=0.9$ and $T=10\mathrm{K}$ is shown in the inset of Fig.~\ref{fig:Tc}. The
electron system is far from saturation, the polarization reaches $\sim16\%$ 
for $U/W=70$ and increases slowly for stronger interaction parameters.\\   
\begin{figure}
    \includegraphics[width=0.9\linewidth]{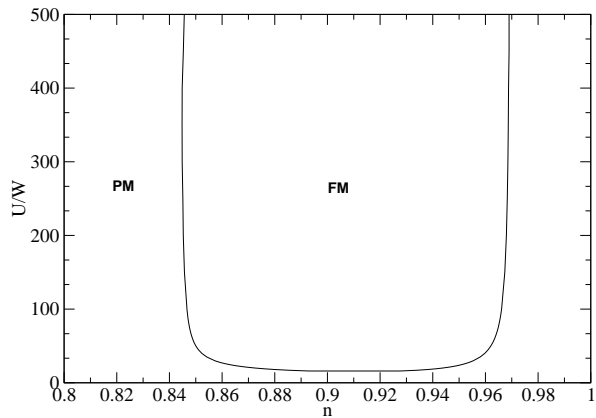}
    \caption{\label{fig:phasediag}Ferromagnetic phase diagram of the 3D simple cubic lattice
    as function of $U/W$ and band filling $n$. Parameters: $T=10\mathrm{K}$.}
\end{figure}
The full phase diagram is shown in Fig.~\ref{fig:phasediag}. The lower phase
boundary is decreasing with increasing $U/W$ to lower $n$ up to $U/W\sim350$ where
it takes the value $n_{crit}\approx0.845$. Increasing $U/W$ further will not
increase the ferromagnetic region but we observe a slight shifting to higher $n$
again. The upper phase boundary growths monotonically with increasing $U/W$
reaching a value of $n\approx0.969$ for $U/W=500$. This result is in accordance
with Nagaokas theorem.\\
As expected, in the local approximation the critical $U$: $U_{crit}/W\approx 5$ 
at $T=10$ K for the 3D SC lattice is lower than in the full $\mathbf{k}$-dependent 
case and agrees well with findings from DMFT calculations ($U/W\gtrsim3$)\cite{Schmitt11}.
\section{\label{sec:summary}summary and conclusions}
In this work we have derived a non-local self-energy for the Hubbard model
within a modified perturbation theory approach.It was shown, that this self-energy 
fulfills a variety of limiting cases (e.g. weak coupling and atomic limit) and shows the 
correct high energy behavior by construction. Numerical tools for the
evaluation of this self-energy were introduced, in particular to solve the
complicated momentum integrations.\\
We show results for the two- and three-dimensional simple cubic lattice,
discussing in detail the influence of non-locality, temperature and interaction
strength on the self-energy, spectral density and quasi-particle density of
states. These results are then used for the interpretation of the calculated
optical conductivity and resistivity curves. \\
The inverse paramagnetic susceptibility was calculated, showing
that there is no ferromagnetic phase transition in the two-dimensional
but for the three-dimensional lattice. The ferromagnetic/paramagnetic phase diagram 
for the three-dimensional lattice is then constructed.\\
Our findings emphasize the importance of non-local correlations in the Hubbard
model in low dimensions, in particular for the fulfillment of Mermin-Wagner
theorem.\\

The strength of the present approach is that it allows self-consistent
calculations at arbitrary band-fillings with a fully $\mathbf{k}$-dependent
self-energy. This comes at the expense of neglecting certain correlation effects. Most
severe in this respect is the missing metal-insulator transition (MIT) at half-filling
which should also occur in finite dimensions at roughly $U/W \approx 1$ as
indicated by DMFT calculations \cite{Fuchs11}. It is interesting to note, that
the (local) MPT, when used as an impurity solver in a DMFT calculation for the infinite
dimensional Hubbard model does show a MIT in the correct $U/W$ region, whereas it
does not when used as a self-energy for the lattice Hamiltonian\cite{Lichtner09}.\\

Another shortcoming of the current state of the theory is the inability to check
for antiferromagnetic phases, which ultimately should appear near half-filling. 
The reason for this is in the Ansatz of the self-energy (\ref{eq:self_mpt}),
which only allows the calculation of homogeneous phases.
This does not mean, that there are no possible antiferromagnetic solutions
within the MPT approach. It would be an interesting task for a forthcoming
work, to extend the MPT in this direction. An indirect hint for the presence of
antiferromagnetic correlations within the MPT can be derived from the breakdown
of ferromagnetic order near half filling, which could have its origin in
competing antiferromagnetic correlations.

\appendix 
\section{coefficients of self-energy expansion}
\begin{eqnarray}
\label{eq:coeff_self1}
	C_{\mathbf{k}\sigma}^{(0)} & = & U\corr{n_{-\sigma}} \\
	C_{\mathbf{k}\sigma}^{(1)} & = & U^2\corr{n_{-\sigma}}(1-\corr{n_{-\sigma}})\nonumber\\
	C_{\mathbf{k}\sigma}^{(2)} & = & C_{\mathbf{k}\sigma}^{(1)}
						             \left(
                                        \epsilon_{\sigma}(\mathbf{k})-\mu+U(1-\corr{n_{-\sigma}})
                                     \right)+U^2B_{\mathbf{k}-\sigma}.\nonumber
\end{eqnarray}
with
\begin{widetext}
\begin{eqnarray}
	B_{\mathbf{k}-\sigma} & = &
			B_{S,-\sigma}+B_{W,-\sigma}(\mathbf{k}) \nonumber													\\
	B_{S,-\sigma} & = & \frac{1}{N}\sum_{i,j}^{i\neq j}T_{ij}\corr{c^{+}_{i-\sigma}c_{j-\sigma}(2n_{i\sigma}-1)}\nonumber	\\
				  & = & \frac{1}{N}\sum_{\mathbf{k}}\left( \epsilon(\mathbf{k})-T_0\right)\int_{-\infty}^{\infty}\mathrm{d}E
				  					f_{-}(E)\left( \frac{2}{U}\left( E-\epsilon_{-\sigma}(\mathbf{k})\right)-1\right)
									S_{\mathbf{k}-\sigma}(E-\mu) \\
	B_{W,-\sigma}(\mathbf{k}) & = & \frac{1}{N}\sum_{i,j}^{i\neq j}T_{ij}\mathrm{e}^{-\mathrm{i}\mathbf{k}(\mathbf{R}_i-\mathbf{R}_j)}
									\left( \corr{n_{i-\sigma}n_{j-\sigma}}-\corr{n_{-\sigma}}
							              -\corr{c^{+}_{j\sigma}c^{+}_{j-\sigma}c_{i-\sigma}c_{i\sigma}}
							  			  -\corr{c^{+}_{j\sigma}c^{+}_{i-\sigma}c_{j-\sigma}c_{i\sigma}}\right).\nonumber
\label{eq:abr1}
\end{eqnarray}
\end{widetext}
\section{high energy expansion of SOPT self-energy}
\begin{equation}
    \Sigma_{\mathbf{k}\sigma}^{(SOC)}(E) \approx
    \sum_{m=1}^{N}\frac{D_{\mathbf{k}\sigma}^{(m)}}{E^m} 
\label{eq:sopt_expand}
\end{equation}
with
\begin{eqnarray}
\label{eq:coeff_self2}
	D_{\mathbf{k}\sigma}^{(1)} & = & U^2\corr{n_{-\sigma}}(1-\corr{n_{-\sigma}}) = C_{\mathbf{k}\sigma}^{(1)}\\
	D_{\mathbf{k}\sigma}^{(2)} & = & U^2\sum_{\mathbf{R}}\mathrm{e}^{\mathrm{i}\mathbf{kR}} \nonumber \\
                               &   &    \left[ \delta_{\mathbf{R,0}}
                                            \left(
                                                \corr{e_{\mathbf{R}-\sigma}}^{(0)}\left(2\corr{n_{\mathbf{R}\sigma}}^{(0)}-1\right)
                                            \right. 
                                        \right. \nonumber \\
                               &   &        \left.
                                                +\corr{n_{\mathbf{R}-\sigma}}^{(0)}\left(M_{\mathbf{R}-\sigma}+M_{\mathbf{R}\sigma}\right)
                                            \right) \nonumber \\
                               &   &    \left.
                                            -\corr{n_{\mathbf{R}-\sigma}}(2M_{\mathbf{R}-\sigma}\corr{n_{\mathbf{R}\sigma}}^{(0)}
                                                                          +M_{\mathbf{R}\sigma}\corr{n_{\mathbf{R}-\sigma}}^{(0)})
                                        \right] \nonumber
\end{eqnarray}
and
\begin{eqnarray}
\corr{n_{\mathbf{R}\sigma}}^{(0)} & = & \int\mathrm{d}Ef_{-}(E)S_{\mathbf{R}\sigma}^{(0)}(E), \\
\corr{e_{\mathbf{R}\sigma}}^{(0)} & = & \int\mathrm{d}E E f_{-}(E)S_{\mathbf{R}\sigma}^{(0)}(E), \\
M_{\mathbf{R}\sigma} & = & T_{\mathbf{R}}-\mu_{\sigma}^{(0)}\delta_{\mathbf{R,0}}.
\end{eqnarray}
\begin{widetext}
\section{\label{sec:optcondderiv} Derivation of the optical conductivity}
The density-density GF in (renomalized: free propagators are replaced by full
ones) diagrammatic one-loop expansion is given by:
\begin{align}
\green{\hat{n}_{\mathbf{k}\sigma}}{\hat{n}_{\mathbf{k'}\sigma}}_{E_n} 
    & \approx 
        \delta_{\mathbf{kk'}}\frac{1}{\beta}\sum_{m}G_{\mathbf{k}\sigma}(iE_{m})
                G_{\mathbf{k}\sigma}(iE_{n}+iE_{m})  \nonumber  \\
    & = \delta_{\mathbf{kk'}}\int\int\mathrm{d}x\mathrm{d}yS_{\mathbf{k}\sigma}(x)
          S_{\mathbf{k}\sigma}(y)\frac{f_{-}(x)-f_{-}(y)}{iE_{n}+x-y}
          \nonumber \\
    & \stackrel{(iE_n\to E+i0^{+})}{=}   
     \delta_{\mathbf{kk'}}\int\mathrm{d}xf_{-}(x)S_{\mathbf{k}\sigma}(x)
         \left( G_{\mathbf{k}\sigma}(x+E+i0^+) 
              + G_{\mathbf{k}\sigma}(x-E-i0^+) 
         \right). \nonumber
\end{align}
From this we get for the correlation part of conductivity (\ref{eq:cond_2}):
\begin{align}
\mathrm{Re}\sigma_{II}^{\beta\alpha}(E+i0^{+}) & = 
    \mathrm{Re}
    \left\{
        \frac{iq^2}{(E+i0^+)}\frac{1}{N}\sum_{\mathbf{k\sigma}}(\partial_{k_{\beta}}\epsilon_{\mathbf{k}})
         (\partial_{k_{\alpha}}\epsilon_{\mathbf{k}}) 
         \int\mathrm{d}xf_{-}(x)S_{\mathbf{k}\sigma}(x)
         \left( G_{\mathbf{k}\sigma}(x+E+i0^+) 
              + G_{\mathbf{k}\sigma}(x-E-i0^+) 
         \right)   
    \right\} \nonumber  \\
& = 
       -2q^2 \delta(E)\frac{1}{N}\sum_{\mathbf{k}\sigma} 
        (\partial_{k_{\beta}}\epsilon_{\mathbf{k}})
        (\partial_{k_{\alpha}}\epsilon_{\mathbf{k}}) 
        \int \mathrm{d}xf_{-}(x)\left(\mathrm{Im}G_{\mathbf{k}\sigma}(x)
         \mathrm{Re}G_{\mathbf{k}\sigma}(x)\right) \nonumber \\
&     +\frac{\pi q^2}{E}\frac{1}{N}
        \sum_{\mathbf{k}\sigma} 
        (\partial_{k_{\beta}}\epsilon_{\mathbf{k}})
        (\partial_{k_{\alpha}}\epsilon_{\mathbf{k}}) 
        \int \mathrm{d}xf_{-}(x)S_{\mathbf{k}\sigma}(x)
        \left(S_{\mathbf{k}\sigma}(x+E)-S_{\mathbf{k}\sigma}(x-E)
        \right).\nonumber
\end{align}
The first term of this result can be recast into the form of the Drude
contribution (\ref{eq:drude_cond}) but with opposite sign:
\begin{align}
-2q^2 \delta(E)\frac{1}{N}\sum_{\mathbf{k}\sigma} 
        (\partial_{k_{\beta}}\epsilon_{\mathbf{k}})
        (\partial_{k_{\alpha}}\epsilon_{\mathbf{k}}) 
&       \int \mathrm{d}xf_{-}(x)\left(\mathrm{Im}G_{\mathbf{k}\sigma}(x)
         \mathrm{Re}G_{\mathbf{k}\sigma}(x)\right) \nonumber \\
& =  -q^2 \delta(E)\frac{1}{N}\sum_{\mathbf{k}\sigma} 
        (\partial_{k_{\beta}}\epsilon_{\mathbf{k}})
        (\partial_{k_{\alpha}}\epsilon_{\mathbf{k}}) 
        \int
        \mathrm{d}xf_{-}(x)\mathrm{Im}\left(G_{\mathbf{k}\sigma}(x)\right)^2
        \nonumber \\
& =   -q^2 \delta(E)\frac{1}{N}\sum_{\mathbf{k}\sigma} 
        (\partial_{k_{\beta}}\epsilon_{\mathbf{k}})
        (\partial_{k_{\alpha}}\epsilon_{\mathbf{k}}) 
        \int
        \mathrm{d}xf_{-}(x)\frac{1}{\partial_{k_{\alpha}}\epsilon_{\mathbf{k}}}
        \partial_{k_{\alpha}}\mathrm{Im}G_{\mathbf{k}\sigma}(x) \nonumber \\
& =   -\pi q^2 \delta(E)\frac{1}{N}\sum_{\mathbf{k}\sigma} 
        (\partial_{k_{\alpha}}\partial_{k_{\beta}}\epsilon_{\mathbf{k}})
        \int
        \mathrm{d}xf_{-}(x)S_{\mathbf{k}\sigma}(x) \nonumber \\
& =   -\delta_{\alpha\beta}\pi\delta(E)q^2 \frac{1}{N}\sum_{\mathbf{k}\sigma} 
        (\partial_{k_{\alpha}}^2\epsilon_{\mathbf{k}})\corr{\hat{n}_{\mathbf{k}\sigma}},\nonumber
\label{Drude2}
\end{align}
where the second and third step is only allowed when the self-energy does not depend on
$\mathbf{k}$ and several steps require a diagonal mass tensor (as it is for the SC
lattice).
\end{widetext}

\end{document}